\documentclass[aps,prc,preprint,groupedaddress,howpacs]{revtex4-1}

\usepackage{graphicx}
\usepackage{epstopdf}
\usepackage{dcolumn}
\usepackage{bm}

\begin{document}


\title{Unbound states of $^{32}$Cl and the $^{31}$S($p$,$\gamma$)$^{32}$Cl
reaction rate} 

\author{M. Mato\v{s}}
\email[E-mail: ]{matos@lsu.edu}
\author{J. C. Blackmon}
\author{L. E. Linhardt}
\affiliation{Department of Physics \& Astronomy, Louisiana State University,
Baton Rouge, LA 70803, USA}

\author{D. W. Bardayan}
\author{C. D. Nesaraja}
\affiliation{Physics Division, Oak Ridge National Laboratory, Oak Ridge, TN
37831, USA}

\author{J. A. Clark}
\affiliation{Physics Division, Argonne National Laboratory, Argonne, IL
60439, USA}
\author{C. M. Deibel}
\affiliation{Physics Division, Argonne National Laboratory, Argonne, IL
60439, USA}
\affiliation{Joint Institute for Nuclear Astrophysics, Michigan State
University, East Lansing, MI 48824, USA}
\author{P. D. O'Malley}
\affiliation{Department of Physics and Astronomy, Rutgers University,
Piscataway, NJ 08854, USA}
\author{P. D. Parker}
\affiliation{Wright Nuclear Structure Laboratory, Yale University, New
Haven, CT 06520, USA}

\date{\today}

\begin{abstract} 
The $^{31}$S($p$,$\gamma$)$^{32}$Cl reaction is expected to provide the
dominant 
break-out path from the SiP cycle in novae and is important for understanding
enrichments of sulfur observed in some nova ejecta. 
We studied the $^{32}$S($^{3}$He,$t$)$^{32}$Cl charge-exchange reaction to 
determine properties of proton-unbound levels in $^{32}$Cl that 
have previously contributed significant uncertainties to the 
$^{31}$S($p$,$\gamma$)$^{32}$Cl reaction rate. 
Measured triton magnetic rigidities were used to determine
excitation energies in $^{32}$Cl.
Proton-branching ratios were obtained by detecting decay 
protons from unbound $^{32}$Cl states in coincidence with tritons. An improved
$^{31}$S($p$,$\gamma$)$^{32}$Cl 
reaction rate was calculated including robust
statistical and systematic uncertainties.
\end{abstract}

\maketitle


\section{Introduction}

A classical nova is a cataclysmic nuclear explosion on the surface of a white dwarf star
resulting from the accretion of hydrogen-rich gas from a companion star. 
A~sequence of nuclear reactions produces a sudden luminosity 
increase by up to a factor of a million
and ejects matter from the white dwarf. 
The time scales of explosive hydrogen burning processes are influenced by
the duration of reaction cycles closed by ($p$,$\alpha$) 
reactions, with break out via ($p$,$\gamma$) reactions competing with
$\beta$-decays. 
The SiP cycle is one such cycle, which is of particular interest for understanding 
novae like 
Nova Her 1991 that are observed to exhibit high sulfur abundances
compared 
to solar values \cite{Wil94,Mat94}.

In the SiP cycle the $^{31}$S($p$,$\gamma$)$^{32}$Cl reaction 
is believed to be the dominant break-out reaction~\cite{Vou94}. 
The rate of the $^{31}$S($p$,$\gamma$)$^{32}$Cl reaction at nova temperatures
is dominated by resonances corresponding to states in the compound nucleus $^{32}$Cl.
The~$^{31}$S($p$,$\gamma$)$^{32}$Cl reaction rate was previously calculated
based on some measured resonance properties and estimates of others based upon the mirror nucleus
$^{32}$P~\cite{Ili99}.  Uncertainties in the rate were 
provided in a subsequent reanalysis~\cite{Ili10}. 
The $^{32}$Cl excitation energies near the proton threshold  of $Q_p$ = 1581.3(6)~keV~\cite{Bha08}
 used by Ref.~\cite{Ili99,Ili10} were based on 
earlier measurements that disagreed at the 2 sigma level 
(by 10-20 keV) \cite{Vou94,Jea89}, and the resonance strengths 
were only constrained based upon properties of states in the mirror.  
A recent study of 5 states below 2.3~MeV via the 
$^{32}$S($^{3}$He,$t$)$^{32}$Cl reaction \cite{Wr210} is in agreement with the excitation 
energies reported by Ref. \cite{Vou94}. 
However, significant uncertainties remain regarding resonance strengths and size of
systematic uncertainties in the resonance energies.

We have studied proton-unbound states
in $^{32}$Cl using the~$^{32}$S($^{3}$He,$t$)$^{32}$Cl charge-exchange reaction.
Excitation energies and proton-branching ratios for states of astrophysical interest for the 
$^{31}$S($p$,$\gamma$)$^{32}$Cl reaction rate were determined.  In the following sections we 
describe the experiment and new results for states in $^{32}$Cl, including the observation of 
a~predicted, but previously unobserved, level.  We then present calculations of a 
new recommended $^{31}$S($p$,$\gamma$)$^{32}$Cl reaction rate 
based on these new results
including  statistical and systematic uncertainties.


\section{Experiment}

\begin{figure*}
\begin{center}
\includegraphics[width=0.9\linewidth]{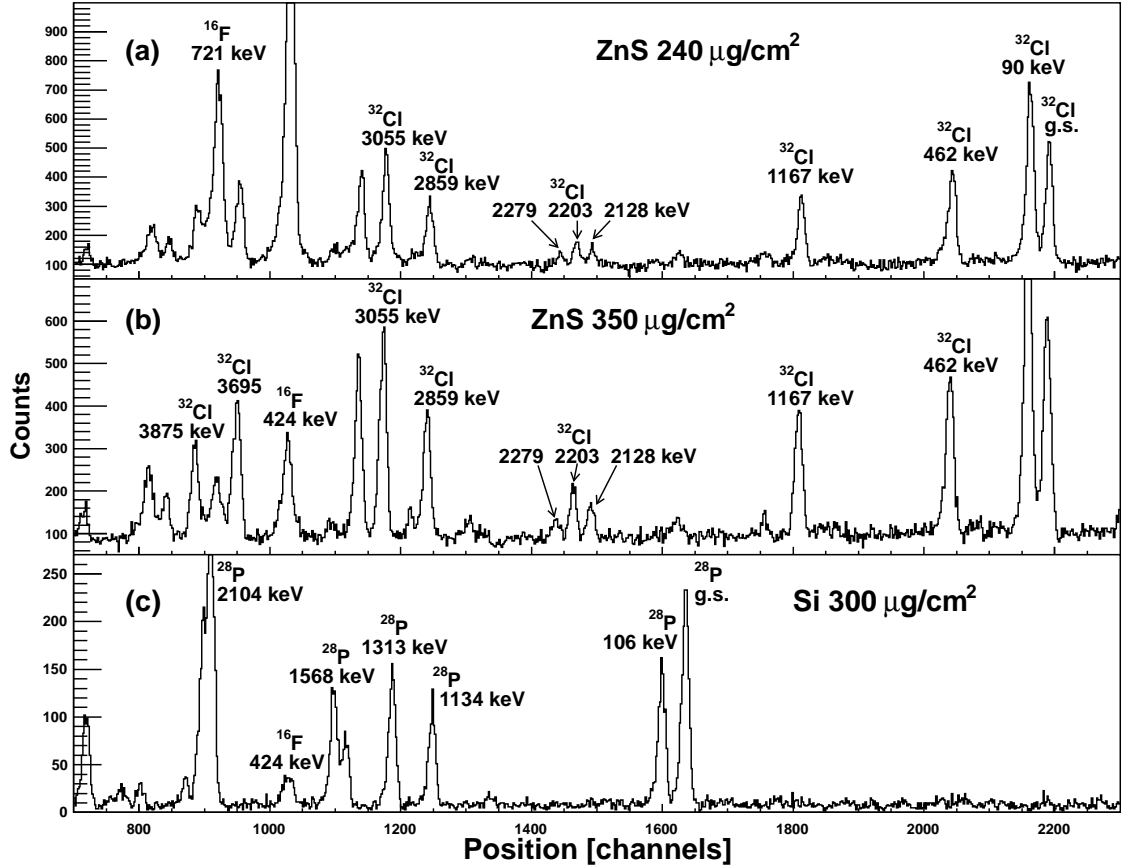}
\caption{Spectra showing position of tritons at the focal plane measured by
the ionization chamber
with the Enge spectrograph set at 3$^{\circ}$. Three spectra show results from
measurements with three
targets, 240 $\mu$g/cm$^2$ ZnS, 350 $\mu$g/cm$^2$ ZnS and 
a~300~$\mu$g/cm$^2$ Si target.
Peaks are identified with the final level in the nucleus produced.
}
\label{f:spectra}
\end{center} 
\end{figure*}

We used the charge-exchange reaction $^{32}$S($^{3}$He,$t$)$^{32}$Cl to
populate states in $^{32}$Cl.
A~30-MeV $^{3}$He$^{2+}$ beam from the Extended Stretched
TransUranium (ESTU) tandem Van de Graaff accelerator at the Wright Nuclear
Structure Laboratory (WNSL) at Yale University bombarded ZnS targets. Targets with thicknesses  
of 240 $\mu$g/cm$^2$ and 350 $\mu$g/cm$^2$ both on 5 $\mu$g/cm$^2$ carbon
substrates were produced via
evaporation at Oak Ridge National Laboratory and used
in the experiment at the WNSL. 
Their thicknesses were determined with about 10\% uncertainty
via energy loss measurements of $\alpha$ particles from a $^{241}$Am
source.
A~150 $\mu$g/cm$^2$ CdS target on a 20 $\mu$g/cm$^2$ substrate was used 
at one angle (5$^{\circ}$) but 
had worse energy resolution and lower 
counting rates and was therefore not used at other angles.
Additional data were taken with a 300 $\mu$g/cm$^2$ 
natural Si target for calibration and a~900 $\mu$g/cm$^2$  Zn target for background analysis.

%

Reaction products were separated and analyzed with the Enge split-pole spectrograph at the WNSL.
The~spectrograph separates particles according to their magnetic
rigidities, $B \rho$, so that discrete positions at the focal plane correspond to
discrete momenta.
The positions of detected particles at an Enge focal plane were determined by 
a~position-sensitive ionization drift chamber filled with
150 Torr of isobutane gas~\cite{Par06}. 
As the ions pass through the gas, the amount of charge collected by the cathode 
determines the energy lost
in the detector. 
The residual energy was measured by a thick plastic
scintillator located behind the ionization chamber.
Measurements were conducted at spectrograph (laboratory) 
angles of 3$^{\circ}$, 5$^{\circ}$,
10$^{\circ}$
and 20$^{\circ}$.
At the 3$^{\circ}$ setting the protons emitted from excited
states in $^{32}$Cl 
were detected in coincidence with outgoing tritons
by the Yale Lamp Shade Array (YLSA)~\cite{Vis03} consisting of four 16-strip
silicon detectors arranged in a lamp-shade configuration.
The YLSA detectors covered an angular
range of $\theta_{lab}$ = 131$^{\circ}$ to $\theta_{lab}$ = 166$^{\circ}$ 
and were calibrated with $\alpha$ particles 
from a $^{241}$Am source.

\section{Level Energies}
Tritons were identified at the focal plane of the spectrograph using 
the~$E_{res}$ vs. $\Delta
E$ relationship from the ionization chamber and the~scintillator. 
Focal-plane position spectra gated on tritons are shown
in Fig.~\ref{f:spectra}, and states in $^{32}$Cl
populated via the $^{32}$S($^{3}$He,$t$)$^{32}$Cl charge-exchange reaction
are labeled. Tritons corresponding to states in 
 $^{16}$F resulting from the $^{16}$O($^{3}$He,$t$)$^{16}$F
 reaction due to oxygen contamination in the targets are also seen in Fig.~\ref{f:spectra}.
 
Positions at the focal plane were calibrated using known states
in $^{28}$P populated from the $^{28}$Si($^{3}$He,$t$)$^{28}$P reaction on a silicon target.
The 424 keV $^{16}$F state populated from oxygen contaminants in the targets, and the ground and 90 keV
first excited states in $^{32}$Cl were also used to calibrate the position spectra, providing 14 calibration 
points spanning the region of interest.
Centroids were obtained by fitting a Gaussian function to each peak with a linear background, and the 
centroids of the calibration peaks were 
fit by their magnetic rigidities. Both 2$^{nd}$ and 3$^{rd}$
order polynomial functions were fit.  Excitation energies for all states of astrophysical
interest agreed within 1 keV from both fits, but the 3$^{rd}$ order fit function was adopted as it
provided a better fit to the lowest and highest energy calibration peaks.
The magnetic rigidities of the recoiling tritons were 
calculated from the reaction kinematics 
with the newest atomic mass values implemented~\cite{AW03,nucmas},
including new values for $^{28}$P and $^{32}$Cl obtained 
recently, with mass excess of $-7147.5(12)$~keV, $-13333.8(12)$ keV, respectively~\cite{Wre10}. 
Small corrections 
($<$30~keV) were made for the energy loss of the incident $^{3}$He ions and
recoiling tritons 
in the target using the energy loss code STOPIT ~\cite{stopit}, assuming
the reaction happened in the center of the target.
The fit function was then used to obtain magnetic rigidities from the centroids for the peaks
of interest. The energies of the levels were calculated from reaction kinematics
taking energy losses into account.
The statistical uncertainty in the energy of each state, $\Delta E_{i}$, was calculated as a combination of the
uncertainties originating from the centroid determination
and the fitting function uncertainties estimated from covariance analysis.
The energy resolution for each triton peak was typically about 40~keV (FWHM), and no peaks
wider than the instrumental resolution were observed.

Excitation energies were obtained for each level at several
angles and with the different targets. 
The final weighted average for each state was calculated from these energies.
The uncertainty in the excitation energy was assigned as  
the larger of the average uncertainty
%
%
\begin{equation}
  \Delta E_{av} = \sqrt{ \frac{\displaystyle 1}
  {\displaystyle \sum_{i=1}^n \frac{\displaystyle 1}{\displaystyle (\Delta
E_i)^2}} }
\end{equation}
and the scatter uncertainty
\begin{equation}
  \Delta E_{scat} = \sqrt{ 
  \frac{1}{n-1} \frac{\displaystyle \sum_{i=1}^n \frac{(E -
E_i)^2}{\displaystyle (\Delta E_i)^2} }
  {\displaystyle \sum_{i=1}^n \frac{\displaystyle 1}{\displaystyle (\Delta
E_i)^2}}
  } .
\end{equation}

Systematic uncertainties were estimated using a Monte Carlo simulation,
where the target thicknesses and mass values were randomly varied with Gaussian
distribution probabilities. Mass and thickness uncertainties were used as standard deviations, $\sigma$.
The systematic error was determined as the standard deviation 
of the Gaussian fit for the level energy distributions obtained in the Monte Carlo 
simulation. The systematic uncertainty was found to be 4~keV for all states,
except for the 462 keV and 1167 keV states, where the systematic uncertainty was 
determined to be 2~keV and 3~keV, respectively.

The excitation energies determined from this measurement with statistical uncertainties
 are given in Table~\ref{t:results}.
They are compared with other measurements \cite{Jea89, Vou94,Wr210}
and the evaluation by \cite{End98}. 
The evaluation favored \cite{Vou94} over \cite{Jea89}, 
where results from both measurements were available
and generally adds 11~keV to \cite{Jea89} if only this one is available. 
The level energies determined in $^{32}$Ar($\beta ^+$p)
experiments include 461.1(1)~keV, 1168.5(2)~keV, and 4076(10)~keV from~\cite{Bjo85} 
and 4072(9)~keV from~\cite{Sch93},
all agreeing well with our results.
A recent measurement also using 
the $^{32}$S($^{3}$He,$t$)$^{32}$Cl reaction provides energies for five levels between 1.3 and 2.3 MeV
that are systematically higher than the current work by about 4 keV ~\cite{Wr210}.


%
 \begin{table}[b]
 \caption{\label{t:results} Excitation energies in $^{32}$Cl measured in this work compared to other
 measurements and evaluations. All listed uncertainties are statistical. 
 Systematic uncertainties have been estimated in this work as 4~keV for all states,
 except for the 462 keV state with the systematic uncertainty of 2~keV and the 1167 keV state with the systematic uncertainty of 3~keV.}
 \setlength{\extrarowheight}{1.3pt}
 \begin{tabular}{ccccc}
\hline
this  &  Jeanperrin & Vouzoukas  & Endt & Wrede\\	
work   &  \cite{Jea89} & \cite{Vou94}  & \cite{End98} & \cite{Wr210} \\
  \ [keV]  & [keV] & [keV] & [keV] & [keV] \\
\hline
462.0(10) &  447(7) &   & 461.1(1) & \\ 
1167.0(21) &  1157(5) &    & 1168.5(2) &  \\ 
1327.4(29) & 1326(5) & 1329(3)  & 1331(3) & 1331.2(5) \\ 
1734.2(14) & 1719(4) & 1735(3) &  1733(2) & 1736.7(6) \\ 
2127.5(19) &  2122(5) & 2129(3)  & 2130(3) & 2131.1(4) \\ 
2203.1(28) &  2193(7) & 2213(3) & 2212(3) & 2209.5(5) \\ 
2278.6(25) &  2270(5) & 2281(3)  & 2281(3) & 2283.5(5) \\ 
2610.9(30) &   &    &  &  \\ 
2677.0(16) &  2665(10) &   &  2676(10)  & \\ 
2859.2(14) &  2858(5) &   &  2869(5)  & \\ 
2931.5(17) &  2941(5) &    &  2952(5)  & \\ 
3054.7(14) & 3056(5) &    &  3067(5)  & \\ 
3163.9(11) &  3166(5) &    &  3177(5)  & \\ 
3280.8(23) &  3290(10) &    &  3301(10)  & \\ 
3695.0(9) &  3692(7) &  &    3703(7)  & \\ 
3874.8(17) &  3883(5) &    &  3894(5)  & \\ 
3999.5(12) &  4002(6) &    &  4013(6)  & \\ 
4073.6(11) &  4080(7) &    &  4074(7)  & \\ 
4349.9(23) &  4356(7) &    &  4367(7)  & \\ 
4577.1(30) &  4590(8) &    &  4601(8)  & \\ 
\hline
 \end{tabular}
 \end{table}

\begin{figure}[tb]
\begin{center}
\includegraphics[width=0.9\linewidth]{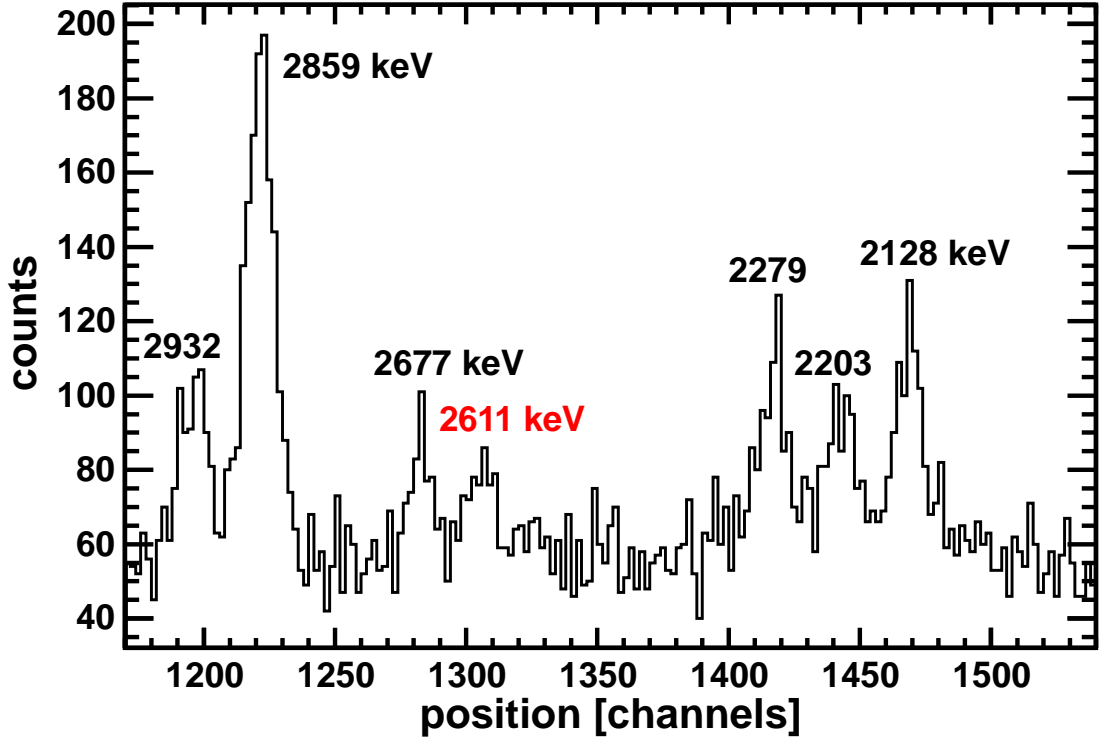}
\caption{(Color online) Focal-plane position spectrum with the Enge spectrograph set at 10$^{\circ}$ showing the newly discovered 2611 keV state
in $^{32}$Cl produced in the 350 $\mu$g/cm$^2$ ZnS target.}
\label{f:newsbranch}
\end{center} 
\end{figure}

We have also discovered an excited state in $^{32}$Cl at  2611(5)~keV,
shown in Fig.~\ref{f:newsbranch}.  The measured excitation energy for the new state is in good 
agreement with a $1^{+}$ state predicted  at 2574(50)~keV  based on the mirror nucleus  $^{32}$P
and estimated using the IMME equation~\cite{Ili99}.

\section{Proton Unbound States}

\begin{figure}[b]
\begin{center}
\includegraphics[width=1\linewidth]{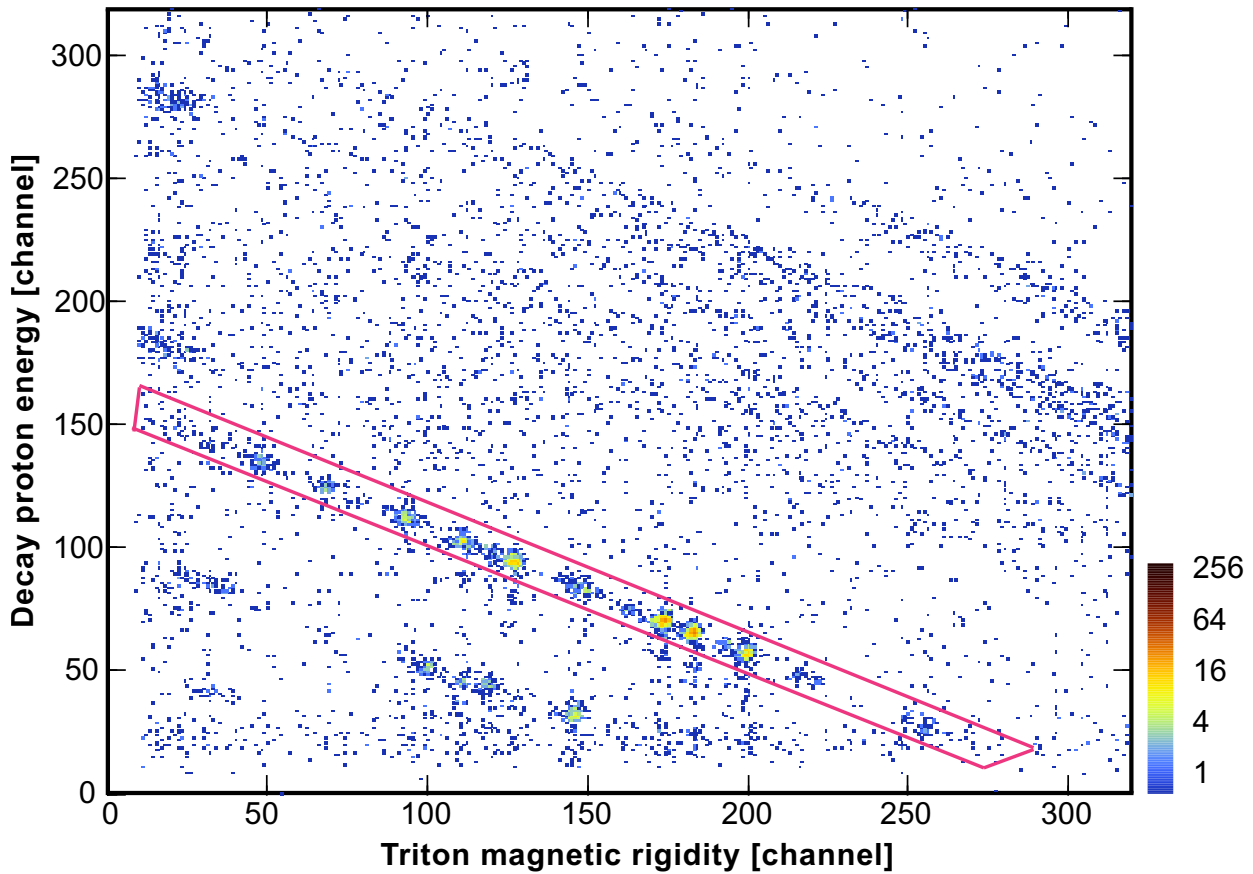}
\caption{(Color online) Triton energy - proton energy coincidence 	spectrum. 
The gated band corresponds to the~$^{32}$S($^{3}$He,$t$)$^{32}$Cl$^*$($p$)$^{31}$S reaction.}
\label{f:tritonproton}
\end{center} 
\end{figure}

The YLSA silicon-detector array was installed covering backward angles at the target position to measure
decays
of proton-unbound states in $^{32}$Cl in coincidence with tritons detected with 
the spectrograph at  $\theta_{lab}$ = 3$^{\circ}$.
Figure \ref{f:tritonproton} shows the energy of the protons detected by YLSA versus
the position (corresponding to momentum) of the tritons at the focal plane.
A gate is shown around events corresponding
to the proton decay of $^{32}$Cl$^{*}$ to the ground state of
$^{31}$S ($J^{\pi} = 1/2^+$).
Proton decay to excited states of 
 $^{31}$S was not possible for $^{32}$Cl excited states below 2.8~MeV due to the 
$^{32}$Cl proton-separation energy of $Q_p$ = 1581.3(6)~keV~\cite{Bha08} and 
the first excited state in $^{31}$S being at 1248.9(2)~keV~\cite{End90}.
The  events below the gated region in Fig.~\ref{f:tritonproton} corresponds 
to the proton decay of $^{16}$F$^{*}$, while
events above are caused by leakage of deuterons into the triton
window in the particle identification cut.

\begin{figure*}[tb]
\begin{center}
\includegraphics[width=0.75\linewidth]{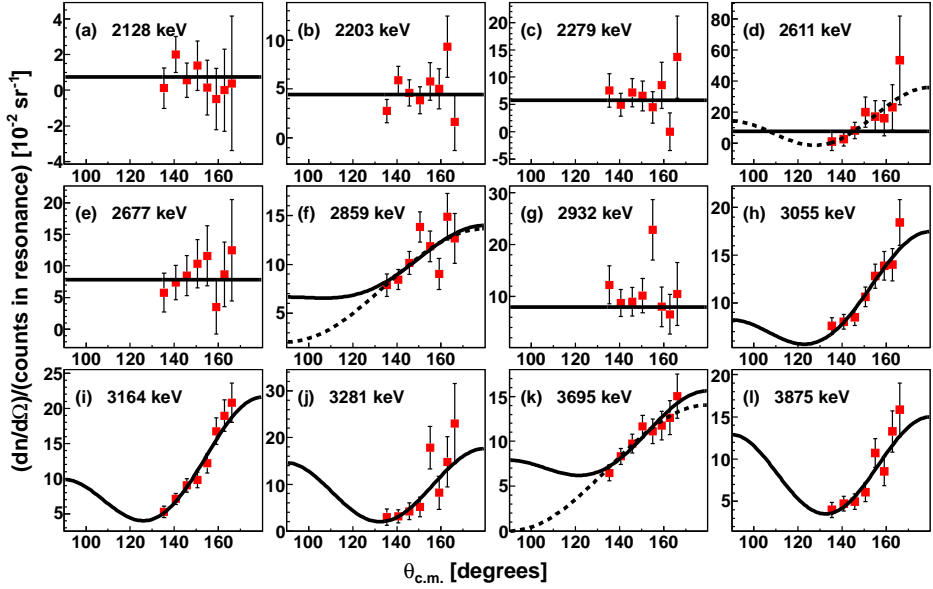}
\caption{(Color online) Triton-proton angular correlation probabilities from the $^{32}$S($^{3}$He,$t$)$^{32}$Cl$^∗$($p$)$^{31}$S reaction at 30 MeV
for the various states in $^{32}$Cl as listed in Table~\ref{t:results}. Red squares are the experimentally determined values with uncertainties,
the black lines are the fits with even Legendre-polynomial terms. The polynomial orders are listed in Table~\ref{t:protons}.
Although the 2$^{nd}$ order
Legendre polynomial (dashed line) for the 2859 keV and 3659 keV states describes the data well, we assume the 4$^{th}$ order
Legendre polynomial to be correct (solid line).}
\label{f:protonangular}
\end{center} 
\end{figure*}

The angular probability distributions of the emitted protons for the $^{32}$Cl
states between 2.1 and 3.9 MeV are shown in Fig.~\ref{f:protonangular}.
Two neighboring YLSA strips are coupled together to reduce scatter due to the low statistics.
The values come from the ratios of the number of events in the $t$-$p$ coincidence peaks to the total
number of tritons populating a given state in $^{32}$Cl, both with background subtracted.
These numbers were then divided by the YLSA efficiency estimated via the use of a Monte Carlo model~\cite{Vis03}.


Angular correlations of isolated nuclear levels  
can be described by a linear combination of even terms of
the Legendre polynomials,
$P_{2k}\left[\cos ( \theta_{c.m.})\right]$, of center-mass angle $\theta_{c.m.}$ up 
to two times the proton
orbital angular momentum $\ell$, i.e.
\begin{equation}
	W(\theta) = \sum_{k=0}^{\ell} A_k P_{2k} \left[\cos (\theta_{c.m.})\right],
	\label{eq:angular}
\end{equation}
which is symmetric around $\theta_{c.m.} = 90^{\circ}$.
The minimum order of the Legendre polynomial needed to fit the data
is defined by statistical significance testing with a p-value~\cite{Wal02}
required to be $>$ 0.05, corresponding to $\chi^2 < $ 14.07,  12.59, and
11.07 for 7, 6, and 5 free parameters, respectively.
This function was constrained to be positive at each point, and
the integral over the full solid angle must be $\le 1$.
The function was then integrated over the full solid angle
to obtain a total proton-branching ratio, $b_p$, as
\begin{equation}
  b_p = \int_{\theta=0}^{\pi} 2 \pi sin(\theta) W(\theta) d\theta.
\end{equation}
The results are shown in Table~\ref{t:protons} with $P_{fit}$ being
the order of the Legendre polynomial used in the fit.
The quoted proton branching-ratio uncertainties come from the uncertainties in~the~fit parameters. 

The proton-branching ratio for the first excited state above the proton separation energy,
$E_{x}$ = 1734.2~keV, is expected to be small. However, even if it was a significant branch, the 
proton energy resulting from the $E_{r}$ = 153 keV resonance
is below the YLSA detector threshold, and would not be observed.
The angular distributions of states with excitation energies of 2128~keV, 2203~keV,
2279~keV, 2611~keV, 2677~keV, and 2932~keV are well fit with just a
constant function, requiring only the first term in Eq.~\ref{eq:angular}. 

The new state at 2611~keV presents an interesting case.  Fits are shown in 
Fig. \ref{f:protonangular} using only an isotropic term (solid line) and the 4$^{th}$ order 
Legendre polynomial (dashed line). 
The angular distribution for this state is better fit with the 4$^{th}$ order
Legendre polynomial, but the 
isotropic fit fulfills the p-value test~\cite{Wal02} at the 95\% confidence level and cannot be ruled out.
As the excitation energy agrees well with that expected for a state corresponding to a 
1$^{+}$ state in the mirror \cite{Ili99}, we adopt the 1$^{+}$ assignment and use 
the isotropic angular distribution required from a $\ell$=0 proton orbital angular momentum.
\begin{table}[b]
\caption{\label{t:protons} Proton-branching ratios for states in $^{32}$Cl. Spin-parities
assigned based on~\cite{Ili99}, except for 3695 keV and 3875 keV states that we assigned based on the mirror symmetry.}
\setlength{\extrarowheight}{1.3pt}
\begin{tabular}{cccc}
\hline
$E_x$ [keV]  & $J^{\pi}$ & $P_{fit}$ & $b_p$ [\%]   \\
\hline
2128 & 3$^+$ & 0 & 7 $\pm$  4    \\
2203 & 1$^+$ & 0 & 54 $\pm$  7   \\
2279 & 2$^+$ & 0 & 66 $\pm$  13    \\
2611 & 1$^+$ & 0 & $>$62\footnote{Fit result is (95$\pm$32)\%.}      \\
2677 & 2$^+$ & 0 & $>$78\footnote{Fit result is (94$\pm$16)\%.}    \\
2859 & 3$^+$ & 4 & $>$95     \\
2932 & 2$^-$ & 0 & $>$88     \\
3055 & 4$^-$ & 4 & $>$97     \\
3164 & 3$^-$ & 4 & $>$96      \\
3281 & 2$^+$ & 4 & $>$88      \\
3695 & 2$^+$ & 4 & $>$96      \\
3875 & 3$^+$ & 4 & $>$94      \\
\hline
\end{tabular}
\end{table}
\begin{table}
\caption{\label{t:strength}
Properties of proton-unbound states in $^{32}$Cl and corresponding resonances
relevant for the $^{31}$S($p$,$\gamma$)$^{32}$Cl reaction.
Energies (including systematic and statistical uncertainties) are the result from this work, except for 3767(10)~keV from~\cite{Bjo85}
and 3397(50)~keV predicted by~\cite{Ili99}. 
The spin-parity assignments from~\cite{Ili99} were used for the states up to 3.5~MeV,
from~\cite{Bjo85} for the 3767 keV state, 
for the 3695 and 3875 keV states were tentatively assigned based on the mirror symmetry.
}
\setlength{\extrarowheight}{1.4pt}
\begin{tabular}{cccccccc}
\hline
 $E_x$  & $E_r$ & $J^\pi$ & $\Gamma_\gamma$ & $\Gamma_{p}$ & $\omega
\gamma$ & $\sigma(\omega \gamma)$ & \\
  \ [keV] & [keV] &  &[meV] & [meV] & [meV] & [meV] \\
 \hline
1734  &  153(5) & 3$^+$ & 1.0 & 2.8$\times$10$^{-8}$ &
4.9$\times$10$^{-8}$  & 1.0$\times$10$^{-8}$ \footnote{The uncertainty distribution of the 153 keV resonance strength includes
an~additional log-normal component with
$\sigma_{log}(\omega \gamma) = 0.58$.}  \\
2128  &  546(5)  & 3$^+$ & 7.9 & 0.59 & 0.96  & 0.61 \\
2203  &  622(5)  & 1$^+$ & 15.5 & 18.2 &  6.3  & 3.6 \\
2279  &  697(5)  & 2$^+$ & 3.1 & 6.0 &  2.54   & 0.76 \\
2611  &  1030(5)  & 1$^+$ & 20.2 & 
		&  14.4  & 7.5  \\
2677  &  1096(5)  & 2$^+$ & 57.9 & 
		&  68 & 30 \\
2859  &  1278(4)  & 3$^+$ & 5.4 &  & 9.5  & 3.7 \\
2932  &  1350(5)  & 2$^-$ & 2.3 &  & 2.84  & 0.76  \\
3055  &  1473(5)  & 4$^-$ & 0.8 &  & 1.81  & 0.42  \\
3164  &  1583(4)  & 3$^-$ & 2 &  &  3.51  & 0.96  \\
3281  &  1700(5)  & 2$^+$ & 15 &  &  18.3 & 8.5  \\
3397 & 1816(50) & 4$^+$ & 1.8 &  & 4.1 & 1.4  \\
3695 & 2114(4) & 2$^+$ & 28 &  &  35 & 20  \\
3767 & 2186(10) & 1$^+$ & 110 &  & 83 & 34  \\
3875 & 2294(4) & 3$^+$ & 59 &  & 104 & 65  \\
\hline
\end{tabular}
\end{table}

For the 2859 keV state, the 2$^{nd}$ order Legendre
polynomial fulfills the
p-value test giving
a total proton-branching ratio of 75$\pm$5 \%. However, a fit with the 4$^{th}$ order
Legendre
polynomial
(dashed line in Fig.~\ref{f:protonangular}) differs from the previous one
only outside
the area covered with our data points, giving a total proton-branching ratio
$>$95\%. The neutron spectroscopic factor for the mirror state in $^{32}$P
has been measured to be 0.03 \cite{Gas73} and 0.008 \cite{Eck89}.  While there
is a discrepancy between the measurements, even the lower value implies an 
expected proton width for the 2859 keV level that would be about 3 orders of 
magnitude larger than the expected gamma width.
Therefore, we adopt the result from the 4$^{th}$ order
Legendre polynomial fit.

The situation is similar for the 3695 keV state. 
We find a total proton-branching ratio
of 59$\pm$4 \% coming from the fit with the 2$^{nd}$ order Legendre
polynomial and a branching ratio of 
$>$96\% using a 4$^{th}$ order
Legendre polynomial.  
While the mirror assignment is not as clear for the 
3695 keV state, the most likely candidate, the 3880 keV ($2^+$), 
has a measured spectroscopic factor of 0.028~\cite{Gas73},
in agreement with $\approx$0.03 predicted for the $^{32}$Cl state by shell model calculations~\cite{Bro11}.
This is a factor of 15 more than the spectroscopic factor required for the branching ratio
of 59\% and we therefore adopt $>$96\% from the 4$^{th}$ order
Legendre polynomial fit.

For the 3055, 3164, 3281 and 3875 keV states, a 4$^{th}$ order
Legendre polynomial fit is required, and the resulting branching ratio is consistent
with $b_p =$ 100\%. 
The lower limit for these states is then statistically estimated based on the number of
events.

The obtained values for minimum proton orbital angular momenta are in a good agreement
with assumed spins and parities. The spin-parity assignments for states with $E_x < 3.5$~MeV 
are taken from~\cite{Ili99}.
The spins of the 3695 and 3875 keV states were tentatively assigned based on the mirror symmetry
corresponding to the 3880.3 keV 2$^+$ and 3989.8 keV $(3^+)$ states in $^{32}$P~\cite{End98},  
as the 3796.1 keV $(1^+)$ in $^{32}$P is assumed to be the mirror state to the 3767 keV
state in $^{32}$Cl that is known to be $J^\pi=1^+$~\cite{Bjo85}.

\section{$^{31}$S(\lowercase{p},$\gamma$)$^{32}$C\lowercase{l} Reaction
Rate}

At nova temperatures, $T\approx$0.1--0.3~GK, the states just above 
the proton-separation energy ($Q_{p}$ = 1581.3~keV)    
dominate the $^{31}$S($p$,$\gamma$)$^{32}$Cl
reaction rate.
As the resonances are generally narrow and well separated, the resonant component
of the reaction rate (in cm$^{3}$mol$^{-1}$s$^{-1}$) can be approximated by
\begin{eqnarray}
  N_A \langle \sigma v \rangle & = & 1.54 \times 10^{11} (\mu T_9)^{-3/2}
\nonumber \\
   & & \times \sum_r (\omega \gamma)_r \exp(-11.605 E_r/T_9),
\label{eq:rate}
\end{eqnarray}
where $T_9$ is the temperature in GK, $E_r$ is the energy of the $^{32}$Cl
resonance in MeV, $\mu$ is the reduced mass in atomic mass units, and $(\omega \gamma)_r$
is
the resonance strength in MeV, given by the spin of the resonance, $J_r$ and its partial
($\Gamma_p$,$\Gamma_{\gamma}$) and total ($\Gamma$) widths  as 
\begin{equation}
	(\omega \gamma)_r = \frac{(2 J_r +1)}{4} \frac{\Gamma_{p} \Gamma_{\gamma}}{\Gamma} .
\label{eq:strength}
\end{equation}
The resonance reaction rate depends exponentially on the resonance energies, $E_r$, and
linearly on the partial widths through the 
resonance strengths, though the proton partial width, $\Gamma_p$, also has an exponential
dependence on energy through the penetrability.
Therefore, our improved measurement of the resonance
energies and the 
proton-branching ratios, corresponding to 
$\Gamma_{p} / \Gamma = \Gamma_{p} / (\Gamma_{p} +
\Gamma_\gamma$), has a direct impact on the uncertainty in the 
$^{31}$S($p$,$\gamma$)$^{32}$Cl reaction rate.

\begin{table}[t]
\caption{\label{t:totalrate} Recommended stellar reaction rates as a function of
the temperature $T$ 
for the reaction $^{31}$S($p$,$\gamma$)$^{32}$Cl.
Lower and upper limits cover the
68.2\% confidence level. 
}
\setlength{\extrarowheight}{-3pt}
\begin{tabular}{cccc}
\hline
\rule{0cm}{0.25cm} Tempe- & Recommended & Low & High \\
rature & rate & rate & rate \\
$T$ &  $N_A \langle \sigma v \rangle$ & $N_A \langle \sigma v \rangle$ & $N_A \langle \sigma v \rangle$ \\
\ [GK] &  [cm$^{3}$mol$^{-1}$s$^{-1}$] & [cm$^{3}$mol$^{-1}$s$^{-1}$] & [cm$^{3}$mol$^{-1}$s$^{-1}$] \\
\hline
\rule{0cm}{0.3cm} 0.01 \rule{0cm}{0.3cm} & 3.68$\times 10^{-44}$ & 2.33$\times 10^{-44}$ & 5.12$\times 10^{-44}$ \\ 
0.015 & 1.84$\times 10^{-37}$ & 1.17$\times 10^{-37}$ & 2.56$\times 10^{-37}$ \\ 
0.02 & 3.08$\times 10^{-33}$ & 1.95$\times 10^{-33}$ & 4.28$\times 10^{-33}$ \\ 
0.03 & 6.33$\times 10^{-28}$ & 4.36$\times 10^{-28}$ & 1.02$\times 10^{-27}$ \\ 
0.04 & 5.39$\times 10^{-23}$ & 1.39$\times 10^{-23}$ & 2.12$\times 10^{-22}$ \\ 
0.05 & 2.69$\times 10^{-19}$ & 8.50$\times 10^{-20}$ & 8.10$\times 10^{-19}$ \\ 
0.06 & 7.56$\times 10^{-17}$ & 2.86$\times 10^{-17}$ & 1.91$\times 10^{-16}$ \\ 
0.07 & 4.10$\times 10^{-15}$ & 1.76$\times 10^{-15}$ & 9.13$\times 10^{-15}$ \\ 
0.08 & 7.98$\times 10^{-14}$ & 3.77$\times 10^{-14}$ & 1.62$\times 10^{-13}$ \\ 
0.09 & 7.86$\times 10^{-13}$ & 4.00$\times 10^{-13}$ & 1.48$\times 10^{-12}$ \\ 
0.1 & 4.82$\times 10^{-12}$ & 2.59$\times 10^{-12}$ & 8.61$\times 10^{-12}$ \\ 
0.15 & 9.73$\times 10^{-10}$ & 6.17$\times 10^{-10}$ & 1.48$\times 10^{-9}$ \\ 
0.2 & 1.22$\times 10^{-8}$ & 8.42$\times 10^{-9}$ & 1.71$\times 10^{-8}$ \\ 
0.3 & 9.86$\times 10^{-7}$ & 5.70$\times 10^{-7}$ & 1.45$\times 10^{-6}$ \\ 
0.4 & 1.41$\times 10^{-4}$ & 7.90$\times 10^{-5}$ & 2.05$\times 10^{-4}$ \\ 
0.5 & 2.99$\times 10^{-3}$ & 1.73$\times 10^{-3}$ & 4.26$\times 10^{-3}$ \\ 
0.6 & 2.27$\times 10^{-2}$ & 1.33$\times 10^{-2}$ & 3.20$\times 10^{-2}$ \\ 
0.7 & 9.47$\times 10^{-2}$ & 5.61$\times 10^{-2}$ & 1.33$\times 10^{-1}$ \\ 
0.8 & 2.73$\times 10^{-1}$ & 1.63$\times 10^{-1}$ & 3.81$\times 10^{-1}$ \\ 
0.9 & 6.17$\times 10^{-1}$ & 3.70$\times 10^{-1}$ & 8.59$\times 10^{-1}$ \\ 
1 & 1.18$\times 10^{0}$ & 7.16$\times 10^{-1}$ & 1.63$\times 10^{0}$ \\ 
1.5 & 8.39$\times 10^{0}$ & 5.59$\times 10^{0}$ & 1.11$\times 10^{1}$ \\ 
2 & 2.37$\times 10^{1}$ & 1.69$\times 10^{1}$ & 3.01$\times 10^{1}$ \\ 
3 & 6.99$\times 10^{1}$ & 5.25$\times 10^{1}$ & 8.71$\times 10^{1}$ \\ 
4 & 1.20$\times 10^{2}$ & 9.08$\times 10^{1}$ & 1.48$\times 10^{2}$ \\ 
5 & 1.63$\times 10^{2}$ & 1.25$\times 10^{2}$ & 2.00$\times 10^{2}$ \\ 
6 & 1.97$\times 10^{2}$ & 1.53$\times 10^{2}$ & 2.41$\times 10^{2}$ \\ 
7 & 2.23$\times 10^{2}$ & 1.75$\times 10^{2}$ & 2.70$\times 10^{2}$ \\ 
8 & 2.41$\times 10^{2}$ & 1.91$\times 10^{2}$ & 2.91$\times 10^{2}$ \\ 
9 & 2.53$\times 10^{2}$ & 2.02$\times 10^{2}$ & 3.06$\times 10^{2}$ \\ 
10 & 2.61$\times 10^{2}$ & 2.08$\times 10^{2}$ & 3.14$\times 10^{2}$ \\ 
\hline
\end{tabular}
\end{table}

We have calculated the gamma widths for the states in $^{32}$Cl using the known half-lives 
of mirror $^{32}$P states, $T_{1/2}$, as well as $\gamma$-branching ratios, $b_{\gamma}$, 
and energies, $E_{\gamma_i}$, of the corresponding transitions from the states \cite{Kan97}. 
Assuming that the reduced transition probabilities,
$B(E_i)$ and $B(M_i)$,
are the same for both mirror nuclei, one can calculate the $\gamma$ width of a state in the mirror 
nucleus as a sum through all possible final state transitions:
\begin{equation}
 \Gamma_\gamma \left(^{32}{\text {Cl}}\right) = \sum_i \frac{E_{\gamma_i}^{(\lambda_i+2)}(^{32}{\text
{Cl}})}{E_{\gamma_i}^{(\lambda_i+2)}(^{32}{\text P})} 
 \frac{b_{\gamma} \hbar \ln(2)}{T_{1/2}(^{32}{\text P})},
\end{equation}
where $\lambda$ is electric or magnetic multipolarity.
This follows a similar prescription as was used in \cite{Ili99,Ili10}.
The lowest possible multipolarities were assumed.
In the case of $M1/E2$ transitions, studies of the mirror nucleus~\cite{Kan97} showed
that $M1$  transitions mostly dominate, and thus $M1$ transitions were adopted 
in the present reaction rate calculations.

For the excited states between 2.1~MeV and 2.3~MeV, where the
proton-branching
ratio was determined to be finite but less than 100\%, we calculated 
the proton widths, $\Gamma_{p}$, directly from the gamma widths with
our measured proton-branching ratios. 
For the higher energies, the resonance strength becomes insensitive to the proton width as
$\Gamma_{p} \gg \Gamma_\gamma$ and $\Gamma_{p} \Gamma_{\gamma}
/ \Gamma
\sim \Gamma_{\gamma}$.

The 1734 keV state, corresponding to the 153 keV resonance, is the one state where the 
proton decay width is important, but where no 
information was extracted from our measurement.  In this case we 
calculated the proton width using the prescription also followed in \cite{Ili99},
\begin{equation}
 \Gamma_{p} 
%
= 2 \frac{\hbar^2}{\mu a_c^2}P_c C^2 S_p \theta_{sp}^2 ,
\end{equation}
%
where $\mu$ is the reduced mass and $a_c = 5.6$~fm is the channel radius. 
The single-particle reduced width, $\theta_{sp}^2 = 0.32$, was
derived from the parameterization for nuclei with the mass number $A$ = 12--50 and bombarding energies $E \le$ 1000~keV
based on optical-model computations and $R$-matrix expressions~\cite{Ili97}.
The penetrability was calculated to be $P_c = 2.9 \times 10^{-15}$.
The spectroscopic factor, $C^2 S_p$, was obtained from the reaction studies
with the mirror nucleus $^{32}$P produced via a neutron transfer,
$^{31}$P($d$,$p$)$^{32}$P, at deuteron energies of 10~MeV \cite{Gas73} and 20~MeV~\cite{Eck89}.
The spectroscopic factors reported in these measurements are discrepant, 0.011 \cite{Gas73} and 0.0054 \cite{Eck89}, and
the average value was adopted in \cite{Ili99,Ili10}. However, we conducted a reanalysis of the experimental
cross section data from  \cite{Gas73} and  \cite{Eck89} using the FRESCO code~\cite{Tho88}.
We find differential cross-sections from both experiments to be best fit with a spectroscopic
factor of 0.011, in agreement with \cite{Gas73}. Thus, we adopt the higher value for the spectroscopic factor and 
from this calculate
the proton width of the 153(5) keV resonance to be 2.8$\times$10$^{-8}$ meV.

The calculated proton and gamma widths, as well as resonance strengths calculated using Eq.~\ref{eq:strength}, are listed
in Table~\ref{t:strength}. The resonance energies determined in this work have been used for the calculations,
except for the 3767 keV state measured in the $^{32}$Ar $\beta$-decay studies \cite{Bjo85} and the 3397 keV state
predicted by~\cite{Ili99} based on the mirror symmetry.

\begin{figure*}[t]
\begin{center}
\includegraphics[width=0.9\linewidth]{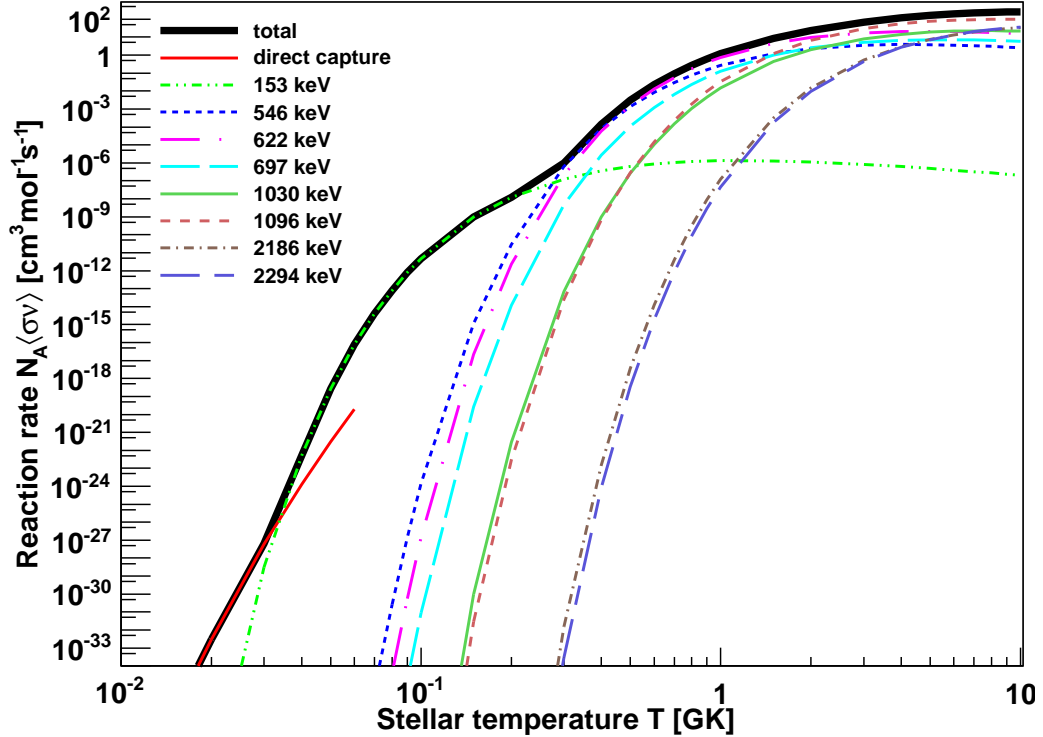}
\caption{(Color online) $^{31}$S($p$,$\gamma$)$^{32}$Cl reaction rate as a function of the stellar temperature $T$.
Resonances with a contribution of
at least 10\% contribution are shown.
}
\label{f:ratecontr}
\end{center} 
\end{figure*}

To cover temperatures below the regions dominated by the resonances, 
we have adopted the direct capture parameterization 
from~\cite{Ili99}. 
The recommended total $^{31}$S($p$,$\gamma$)$^{32}$Cl reaction rate for the stellar temperature range $T$ = 0.01--10 GK
is given in Table~\ref{t:totalrate}. The individual contributions of direct and resonant capture
are illustrated in Fig.~\ref{f:ratecontr}.
The rate has been parameterized in the Reaclib format~\cite{Thi87} 
as the sum of three exponentials, each
with a set of 7 parameters
in the form
\begin{eqnarray}
N_A \langle \sigma v \rangle =  \exp & \left[ a_0 + a_1/T_9 + a_2/T_9^{1/3} + a_3 T_9^{1/3} \right.
\nonumber \\
 &  \left. + a_4 T_9 + a_5 T_9^{5/3} + a_6\times \ln(T_9) \right] 
\end{eqnarray}
where $T_9$ is the temperature in GK, using the tools in~\cite{nucastro}. The fit agrees with the data within 1.5\%.
The new Reaclib parameters are listed in Table~\ref{t:reaclib}.

Uncertainties in the reaction rate have been estimated using a Monte Carlo technique,
as a~combination of normal and log-normal distributions of uncertainties
complicate the analysis.
Uncertainties in the resonance energies contribute to the log-normal distribution,
as the energy is in the exponential of Eq.~\ref{eq:rate}.
The gamma width contributes to the normal distribution,
as its uncertainty originated from the half-life uncertainty, mirror-symmetry
assumption (that we have estimated to be 20\% based on the mirror states in the neighboring nuclei)
and the uncertainty of the proton-branching ratio.
The uncertainty in the proton width will have a pure normal distribution, if extracted
from the proton-branching ratio.  For the 153 keV resonance,
the exponential dependence of 
the penetrability on the energy contributes to the log-normal distribution, but a 
normally distributed contribution also originates in the spectroscopic
factor uncertainty that we have estimated to be 30\%, 
 the spectroscopic factor being different for both mirror nuclei 
due to the effect of Coulomb and other isospin-nonconserving interactions ($\sim$10\%),
a small uncertainty of the wavelength in the penetrability ($\sim$2\%), 
and the reduced width uncertainty that we have estimated to be 20\%.
For the 3695 keV and 3875 keV states, the uncertainty in the spins is considered.
Estimated uncertainties for the resonance strengths are shown in Table~\ref{t:strength}. 
Direct capture uncertainties were taken directly from~\cite{Ili10}.
\begin{table}[t]
\caption{\label{t:reaclib} Recommended Reaclib parameters for the $^{31}$S($p$,$\gamma$)$^{32}$Cl reaction rate
within $T$ = 0.01--10 GK.}
\begin{tabular}{cccccccc}
\hline
set & a$_0$ & a$_1$ & a$_2$ & a$_3$ & a$_4$ & a$_5$ & a$_6$ \\ 
\hline
1 & 297.13 & -2.6702 & 114.07 & -485.67 & 63.95 & -5.699 & 140.32 \\
2 & -35.362 & 4.1263 & -324.5 & 373.08 & -18.087 & 0.9159 & -205.99 \\
3 & 1315.5 & -1.8787 & 330.21 & -1911.9 & 261.17 & -25.381 & 511.02 \\
\hline
\end{tabular}
\end{table}

\begin{figure}[t]
\begin{center}
\includegraphics[height=17cm]{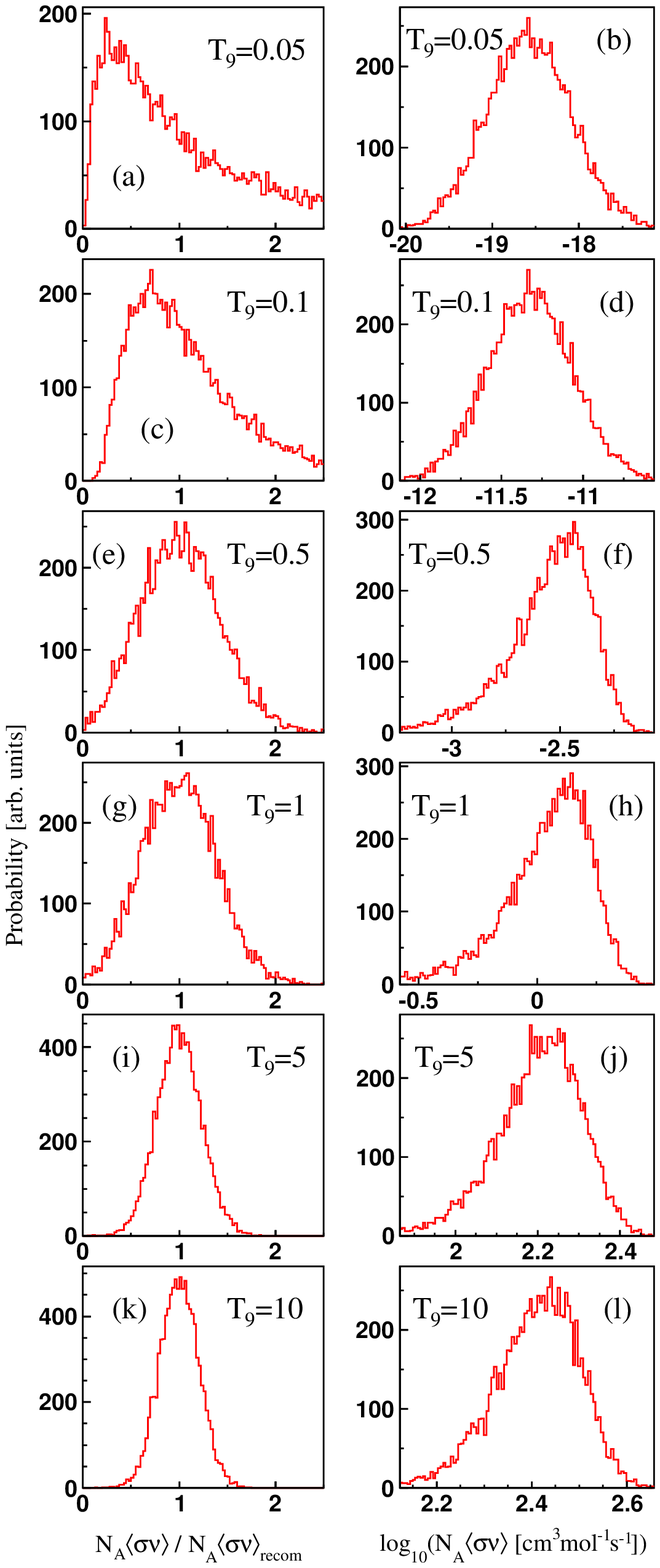}
\caption{(Color online) Total reaction rate probability density functions as 
a~result of the Monte Carlo simulation of the input-uncertainty propagation
for various temperatures. The left figures show the distributions
in the logarithmic scale normalized to the recommended value,
the right figures show the same distributions in logarithmic scale without normalization.
}
\label{f:ratedistr}
\end{center} 
\end{figure}

In the simulation, the values of the energies and resonance strengths were 
varied randomly as a Gaussian distribution. 
Correlations, when dependent on the same parameters, were taken into account.  
The resulting reaction rate distributions (see Fig.~\ref{f:ratedistr}) have various shapes for different temperatures,
including a nearly pure log-normal distribution at $T$ = 0.05~GK and normal distribution at $T$ = 5~GK.
To give final uncertainties that would correspond to the standard deviation $\sigma$ in both
distributions, we found a lower limit as a value with a percentile of 15.9 and 
an upper limit with a percentile of 84.1, covering the
68.2\% confidence level. The results from the Monte Carlo simulation
are listed as the low and high rates in Table~\ref{t:totalrate}.

\section{Discussion}

The total $^{31}$S($p$,$\gamma$)$^{32}$Cl reaction rate and individual contributions 
based upon this work are illustrated in Fig.~\ref{f:ratecontr}.
Direct capture dominates the reaction rate up to $T \sim $ 0.03~GK.
The 153 keV, 546 keV, 622 keV, and 1096 keV resonances, corresponding
 to 1734 keV, 2128 keV, 2203 keV, and 2677 keV levels 
in $^{32}$Cl, dominate the rate over nearly all temperatures.  The
697 keV resonance contributes more than 10\% at nova temperatures, and the
1030 keV resonance must be considered at X-ray burst temperatures, $T\approx 2$~GK.
The 2186 keV, and 2294 keV resonances do not contribute except at very high temperatures,
$T > 5$~GK, and the 2859 keV and 3695 keV levels (which had some
ambiguity as to the shape of the proton angular distribution) 
make a negligible contribution to the reaction rate.

\begin{figure}[t]
\begin{center}
\includegraphics[width=1\linewidth]{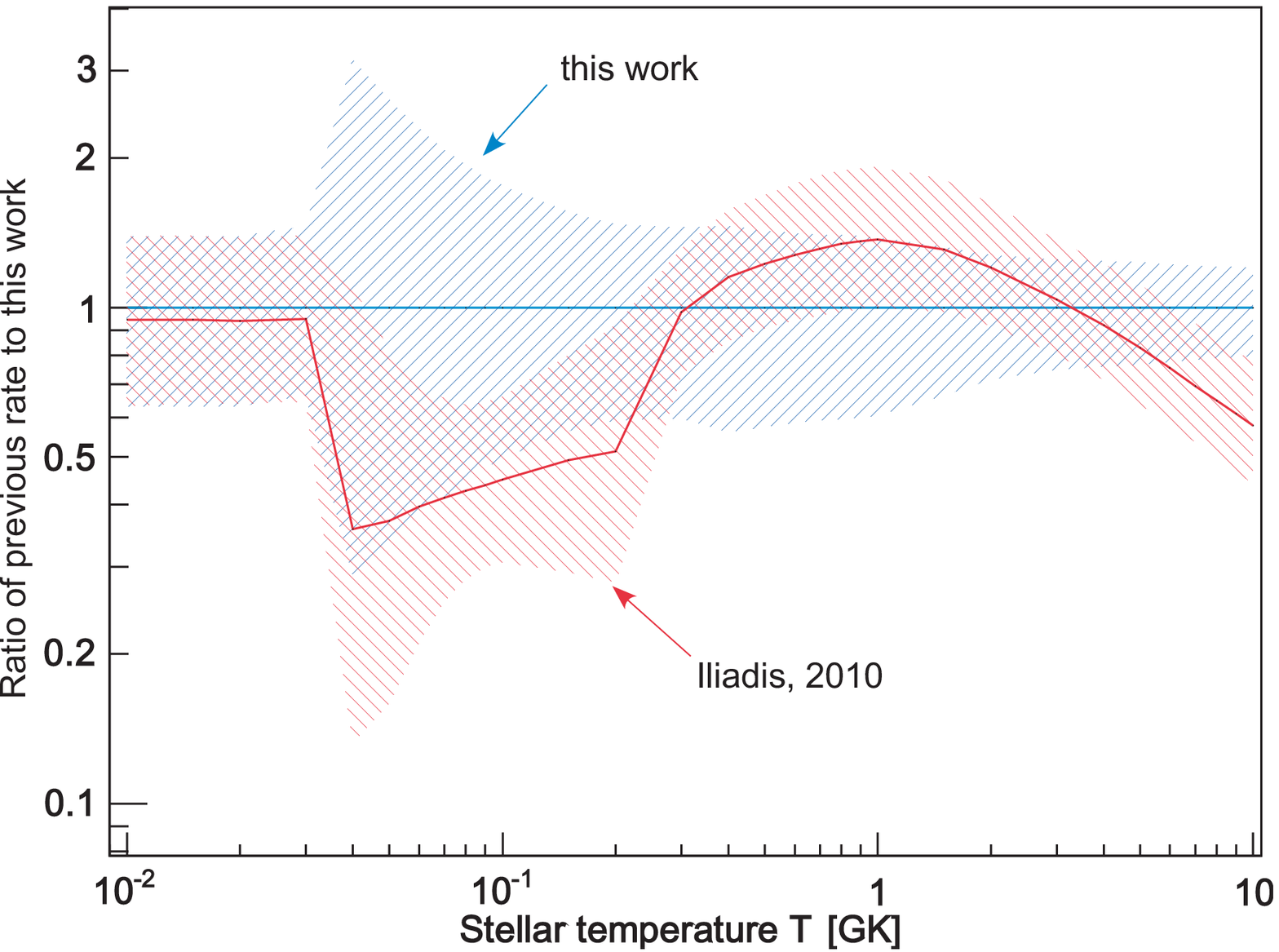}
\caption{(Color online) The $^{31}$S($p$,$\gamma$)$^{32}$Cl reaction rate with 
uncertainties calculated by~\cite{Ili10} (red), with only statistical uncertainties considered, 
is compared to the results
from this work (blue), with both statistical and systematic uncertainties considered. 
Values are shown normalized to the recommended rate from this work. }
\label{f:ratecomp}
\end{center} 
\end{figure}

The $^{31}$S($p$,$\gamma$)$^{32}$Cl reaction rate with uncertainties was recently calculated 
by \cite{Ili10} using the previous work of~\cite{Ili99} with the evaluated level energies from~\cite{End98}.
The ratio of the rate from Ref.~\cite{Ili10} to our rate  is shown in Fig.~\ref{f:ratecomp}.
The uncertainties 
in both rates are illustrated by the hashed regions.  The agreement for $T < 0.03$~GK is expected,  
as the direct capture rate was calculated based on the same parameters~\cite{Ili99},
and the lower and upper limits we take from~\cite{Ili10}.  

\begin{figure}[]
\begin{center}
\includegraphics[width=1\linewidth]{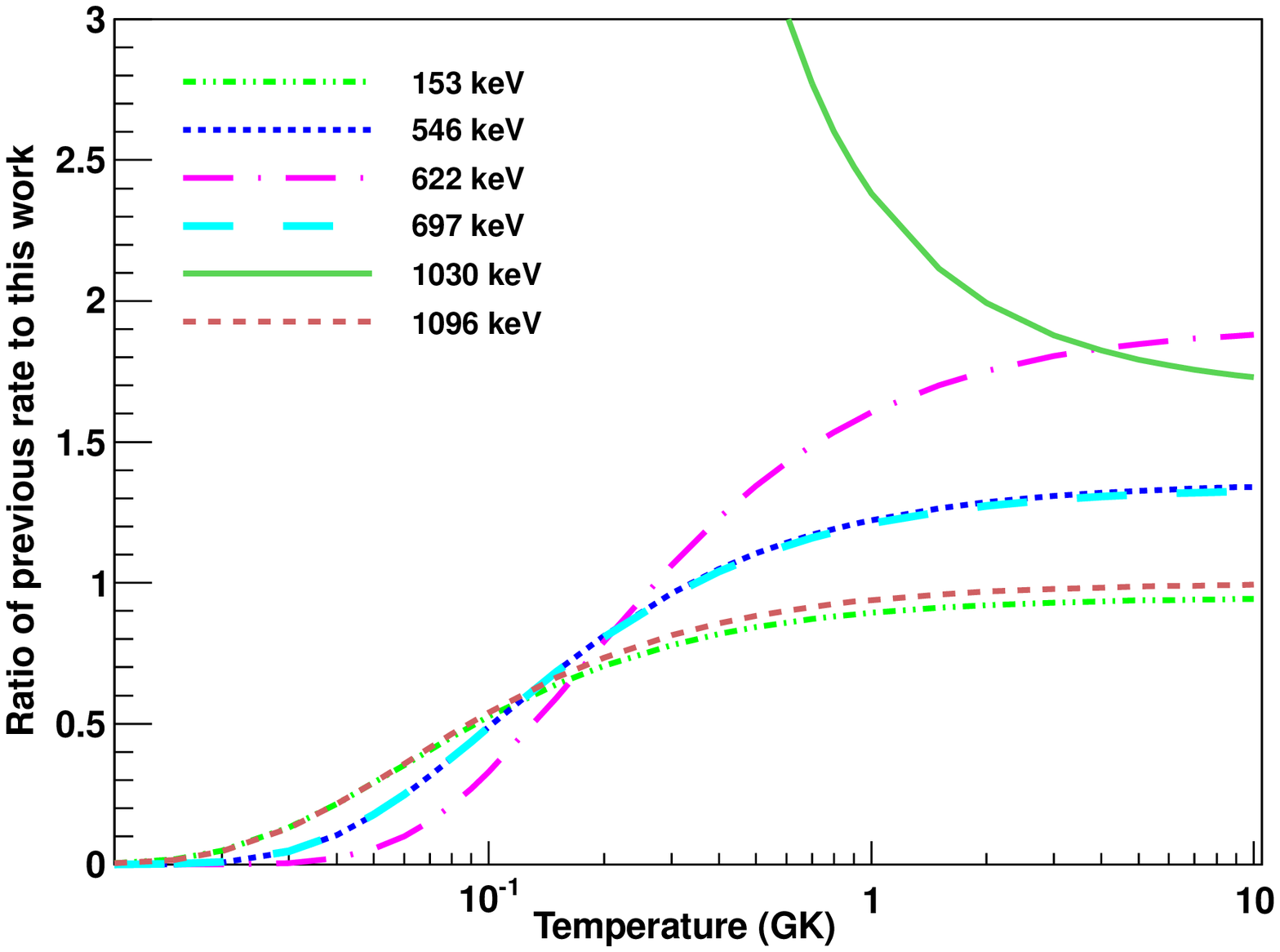}
\caption{(Color online) The ratio of the individual resonance reaction rates 
 from Ref.~\cite{Ili10} to rates from this work as a function of stellar temperature
 for the most important 
resonances contributing to
the $^{31}$S($p$,$\gamma$)$^{32}$Cl rate.}
\label{f:individualres}
\end{center} 
\end{figure}

Over much of the range of nova temperatures 
our recommended rate is significantly greater than even the ``high rate''
recommended in Ref. \cite{Ili10}. This arises from the contribution of individual resonances. 
In Fig.~\ref{f:individualres} the individual resonance reaction rates  from  
Ref.~\cite{Ili10} are compared to our results. 
Our higher reaction rate at most nova temperatures arises 
from the fact that the resonance energies
adopted in Ref. \cite{Ili99} (and derived from \cite{End98}) are greater than 
our energies by 6-15 keV (9 keV on average).
The excitation energies adopted by \cite{End98} and 
\cite{Ili99} primarily reflect a weighted average of \cite{Vou94} and \cite{Jea89} after the results of 
\cite{Jea89} were shifted to match the 1168 keV excitation energy.  The 3-5 keV uncertainties 
in the adopted excitation energies do not properly reflect systematic 
uncertainties, such as uncertainties in the calibration used or the discrepancies between 
measurements.

Our rate also differs from that of \cite{Ili10} due to improved values for resonance strengths.
It should be noted that while we adopt a resonance strength for the 153 keV resonance
that is nearly the same as \cite{Ili99,Ili10}, this arises from two significant changes that
largely cancel each other.  Our lower resonance energy results in a significantly smaller 
penetrability, but we recommend a significantly greater spectroscopic factor based on our
reanalysis of the transfer data of ~\cite{Gas73,Eck89}.  Our reaction rate near peak 
nova temperatures, $T \approx 0.3$~GK, becomes smaller than
that of \cite{Ili10} due to our improved values for the proton branching ratios of the 546 and 622 keV
resonances.  We find $\Gamma_p$ for the 546 keV resonance to be about 30\% smaller than
estimated by \cite{Ili99} and $\Gamma_p$ for the 622 keV resonance to be about 2 orders of magnitude
smaller.  The smaller proton widths result in smaller resonance strengths and reaction rates, though the
magnitude of the effect is mitigated by the lower value of the resonance energies for these states that  
makes for a smaller decrease in the reaction rate than would otherwise be.

At higher temperatures (above about 2 GK) our rate increases in comparison to~\cite{Ili99}
due to our inclusion of resonances above 2~MeV that were not previously considered.

In summary, we have significantly improved the resonance energies and resonance strengths for some of
the most important resonances in the $^{31}$S($p$,$\gamma$)$^{32}$Cl reaction. 
An important aspect of the current work is that we have given careful consideration
to uncertainties, including systematic uncertainties in the level energies, states used for calibration, and
target thickness effects. The largest uncertainties in the reaction rate at nova temperatures arise from the
systematic uncertainty in the resonance energies and the resonance strength of the 153 keV resonance.  
Our excitation energy for the 1734.2(14) keV state (corresponding to the 153 keV resonance) differs from the
value of 1736.7(6) reported by \cite{Wr210} (which used a slightly different set of calibration reactions) 
by 2.5 keV or slightly more than $1\sigma$.  However, 
we estimate the systematic uncertainty in the resonance energies to be 4 keV, which is in agreement with
the fact that the excitation energies for all levels reported by \cite{Wr210} are higher on average 
than this work by about 4 keV.  
Additional experimental information leading to an improvement in the resonance energies would 
therefore be valuable. As the states most important for novae have substantial branches for gamma decay,
an accurate measurement of gamma-ray energies using a complementary approach, for example as 
in \cite{Sew07}, would be particularly helpful in reducing the 
systematic uncertainties that arise largely from Q-value uncertainties in reaction studies like this one.
A direct measurement of the resonance strength of the 153 keV resonance 
(or corresponding proton width) is also desired, but would be experimentally challenging.
 
\section*{Acknowledgments}

The authors would like to thank the staff of the WNSL for their help and support during the measurement.
The authors also thank B.~Alex Brown for his help with the state configurations in $^{32}$P
needed for the DWBA calculation, the estimate of the effect of the~Coulomb 
and  other isospin-nonconserving  interactions on the spectroscopic factor, as well as
the shell-model spectroscopic factor calculation for the 3695~keV state,
and Christian Iliadis for helpful discussions and for providing unpublished information
regarding details of the analysis in \cite{Ili99}. This work is supported by the U.S. Department of Energy 
Office of Nuclear Physics
under Contract No. DE-AC02-06CH11357 and DE-FG02-91ER40609, and grant No. DE-FG02-96ER40978.
 

\end{document}